\documentclass[submitting]{nst}
\usepackage{subfigure,dcolumn}
\usepackage[T2A,T1]{fontenc}
\usepackage[english]{babel}
\usepackage{float}
\usepackage{color}
\usepackage{graphicx}

\usepackage{listings}
\begin{document}

\title{Beam test results of the prototype of the multi wire drift chamber for the CSR external-target experiment}

\thanks{ Supported by the National Natural Science Foundation of China under Grant No. 11927901, the Strategic Priority Research Program of Chinese Academy of Sciences (Grant No. XDB34000000), the National Key R\&D Program of China (Grant No. 2018YFE0205200), the Natural Science Foundation of China (Grant No. 11875301, 11875302, U1867214, U1832105, U1832167), the CAS "Light of West China" Program, and by Tsinghua University Initiative Scientific Research Program. ZQ and ZH  contributed equally.}

\author{Zhi Qin}
\email{qinz18@mails.tsinghua.edu.cn}
\affiliation{Department of Physics, Tsinghua University, Beijing 100084, China}

\author{Zhoubo He}
\affiliation{Institute of Modern Physics, Chinese Academy of Sciences, Lanzhou 730000, China}
\affiliation{School of Nuclear Science and Technology, University of Chinese Academy of Sciences, Beijing 100049, China}

\author{Zhe  Cao}
\affiliation{University of Science and Technology of China, Hefei 230026, China}

\author{Tao Chen}
\affiliation{University of Science and Technology of China, Hefei 230026, China}

\author{Zhi Deng}
\affiliation{Department of Engineering Physics, Tsinghua University, Beijing 100084, China}

\author{Limin Duan}
\affiliation{Institute of Modern Physics, Chinese Academy of Sciences, Lanzhou 730000, China}
\affiliation{School of Nuclear Science and Technology, University of Chinese Academy of Sciences, Beijing 100049, China}

\author{Dong Guo}
\affiliation{Department of Physics, Tsinghua University, Beijing 100084, China}

\author{Rongjiang Hu}
\affiliation{Institute of Modern Physics, Chinese Academy of Sciences, Lanzhou 730000, China}
\affiliation{School of Nuclear Science and Technology, University of Chinese Academy of Sciences, Beijing 100049, China}

\author{Jie Kong}
\affiliation{Institute of Modern Physics, Chinese Academy of Sciences, Lanzhou 730000, China}
\affiliation{School of Nuclear Science and Technology, University of Chinese Academy of Sciences, Beijing 100049, China}

\author{Canwen Liu}
\affiliation{Department of Engineering Physics, Tsinghua University, Beijing 100084, China}

\author{Peng Ma}
\affiliation{Institute of Modern Physics, Chinese Academy of Sciences, Lanzhou 730000, China}
\affiliation{School of Nuclear Science and Technology, University of Chinese Academy of Sciences, Beijing 100049, China}





\author{Xianglun Wei}
\affiliation{Institute of Modern Physics, Chinese Academy of Sciences, Lanzhou 730000, China}
\affiliation{School of Nuclear Science and Technology, University of Chinese Academy of Sciences, Beijing 100049, China}

\author{Shihai Wen}
\affiliation{Institute of Modern Physics, Chinese Academy of Sciences, Lanzhou 730000, China}
\affiliation{School of Nuclear Science and Technology, University of Chinese Academy of Sciences, Beijing 100049, China}

\author{Xiangjie Wen}
\affiliation{Institute of Modern Physics, Chinese Academy of Sciences, Lanzhou 730000, China}
\affiliation{School of Nuclear Science and Technology, University of Chinese Academy of Sciences, Beijing 100049, China}

\author{Junwei Yan}
\affiliation{Institute of Modern Physics, Chinese Academy of Sciences, Lanzhou 730000, China}
\affiliation{School of Nuclear Science and Technology, University of Chinese Academy of Sciences, Beijing 100049, China}

\author{Herun Yang}
\affiliation{Institute of Modern Physics, Chinese Academy of Sciences, Lanzhou 730000, China}
\affiliation{School of Nuclear Science and Technology, University of Chinese Academy of Sciences, Beijing 100049, China}

\author{Zuoqiao Yang}
\affiliation{Institute of Modern Physics, Chinese Academy of Sciences, Lanzhou 730000, China}
\affiliation{School of Nuclear Science and Technology, University of Chinese Academy of Sciences, Beijing 100049, China}

\author{ Yuhong Yu}
\affiliation{Institute of Modern Physics, Chinese Academy of Sciences, Lanzhou 730000, China}
\affiliation{School of Nuclear Science and Technology, University of Chinese Academy of Sciences, Beijing 100049, China}

\author{Zhigang Xiao}
\email{xiaozg@tsinghua.edu.cn}
\affiliation{Department of Physics, Tsinghua University, Beijing 100084, China}

\begin{abstract}
The half-size prototype of the  multi wire drift chamber (MWDC) for the cooling storage ring (CSR) external-target experiment (CEE) was assembled and tested in 350 MeV/u Kr+Fe reactions on the heavy ion research facility in Lanzhou (HIRFL). The prototype consists of 6 sense layers, where the sense wires are stretched in three directions X, U and V, meeting $0^\circ$, $30^\circ$ and $-30^\circ$ with respect to the vertical axis, respectively.  The sensitive area of the prototype is $76 {\rm cm} \times 76 {\rm cm}$. The amplified and shaped signals from the anode wires are digitized in a serial capacity array. Being operated with 1500 V high voltage on the anode wires, the efficiency for each layer is beyond 95\%. The tracking residual is about  $301 \pm 2 \rm \mu m$. The performance meets the requirements of CEE.   

\end{abstract}


\maketitle
\section {Introduction}  \label{sec. I}

Heavy ion collisions (HICs) in the beam energy region from hundreds MeV/u to a few GeV/u  create nuclear matter of a few times of saturation density~\cite{LRP2023,Reis2004,Fuf2008,ZM2009,Star20221,Star20222}. And in return, people can study the nuclear matter equation of  state (nEOS) through the identified observables from HICs, including direct flow, elliptic flow, triangle flow and meson productions etc ~\cite{LBA2014}. Thanks the operations of various experimental devices,  including  KAOS, FOPI, HADES and $\rm S\pi RIT$ as an incomplete list,  great progress has been made in the constraint of nEOS in the last decades ~\cite{Reis2012, Dani2002, Fuchs2001, Sturm2001, Hart2006, HADES2020}. 

The most uncertain part of the nEOS so far is the isovector sector, namely the density dependence of the nuclear symmetry energy $E_{\rm sym}(\rho)$, particularly above $\rho_0$, where $\rho_0$ is the saturation density. Enormous progress has been made in the last decade in constraining the density behavior of $E_{\rm sym}(\rho)$ near  $\rho_0$ by  applying  various experimental probes, including $\pi^-/\pi^+$ yield, neutron-to-proton differential flow etc, in comparison  with the prediction of transport model calculations  ~\cite{Reis2007,XZG2009,XZG2014,WYJ2020,Dan2011,SpRIT2021}.   Despite of the progresses, however, due to the complicated feature of the production and the transport of mesons and light nuclei in HICs, the studies on precise modeling the HIC and the extraction of  $E_{\rm sym}(\rho)$  are still ongoing, known as the transport model evaluation project (TMEP)~\cite{ZYX2018,Ono2019,Colo2021}.  Experimental efforts are ongoing as well. Combining the observation of the gravitation wave from the neutron star merging event GW170817 ~\cite{LIGO2017,LIGO2018}  and the nuclear physics data from terrestrial laboratory, the constraint becomes more convergent~\cite{Huth2022, LYY2021, ZY2019}, but  the uncertainty  remains significant and some questions are still open, like the convolution of the clustering and the transport of isospin degree of freedom~\cite{Yijie2022,Yijie2023}, the inconsistency between the neutron skin thickness of $\rm ^{208}Pb$ and $\rm ^{48}Ca$  ~\cite{PREX2021,Reed2021,CREX2022,Rein2022,ZZ2023} etc.  The constraint of $E_{\rm sym}(\rho)$ becomes unprecedentedly more important  than ever since the observation of GW170817, for the recent review, one can refer to ~\cite{Colo2020, LBA2021, LXF2022}. 

Because of the long standing interest and significance of the studies of nEOS, some new facilities providing heavy ion beams from several hundreds MeV/u to a few GeV/u are constructed or proposed around the world, like CBM at FAIR, MPD at NICA, LAMPS at RAON etc ~\cite{CBM2005,MPD2011,LAMPS2023} . The cooling storage ring (CSR) at the heavy ion facility in Lanzhou (HIRFL) is an accelerator complex, providing  beams of all stable ion species, with the maximum energy of 2.8 GeV/u for proton and 0.5 GeV/u for uranium, respectively \cite{YYJ2020}. If equipped by an advanced experimental device, HIRFL-CSR can provide novel opportunities to the studies on many topics including the properties of nuclear matter. For instance, one can revisit the $\pi^-/\pi^+$ yield ratio, $\rm t/^3He$ yield ratio in HICs at hundreds MeV/u beam energies \cite{ZM2009,GD20241}. Besides,  using the proton induced collisions on heavy targets at 2.8 GeV incident energy, the strangeness baryon $\Lambda$ can be created, and hence one can investigate  the p-$\Lambda$ interactions in cold nuclear matter at normal density via the correlation functions. 

\begin{figure}[!htb]
\centering
\hspace{-0.7cm}
\includegraphics[width=0.5\textwidth]{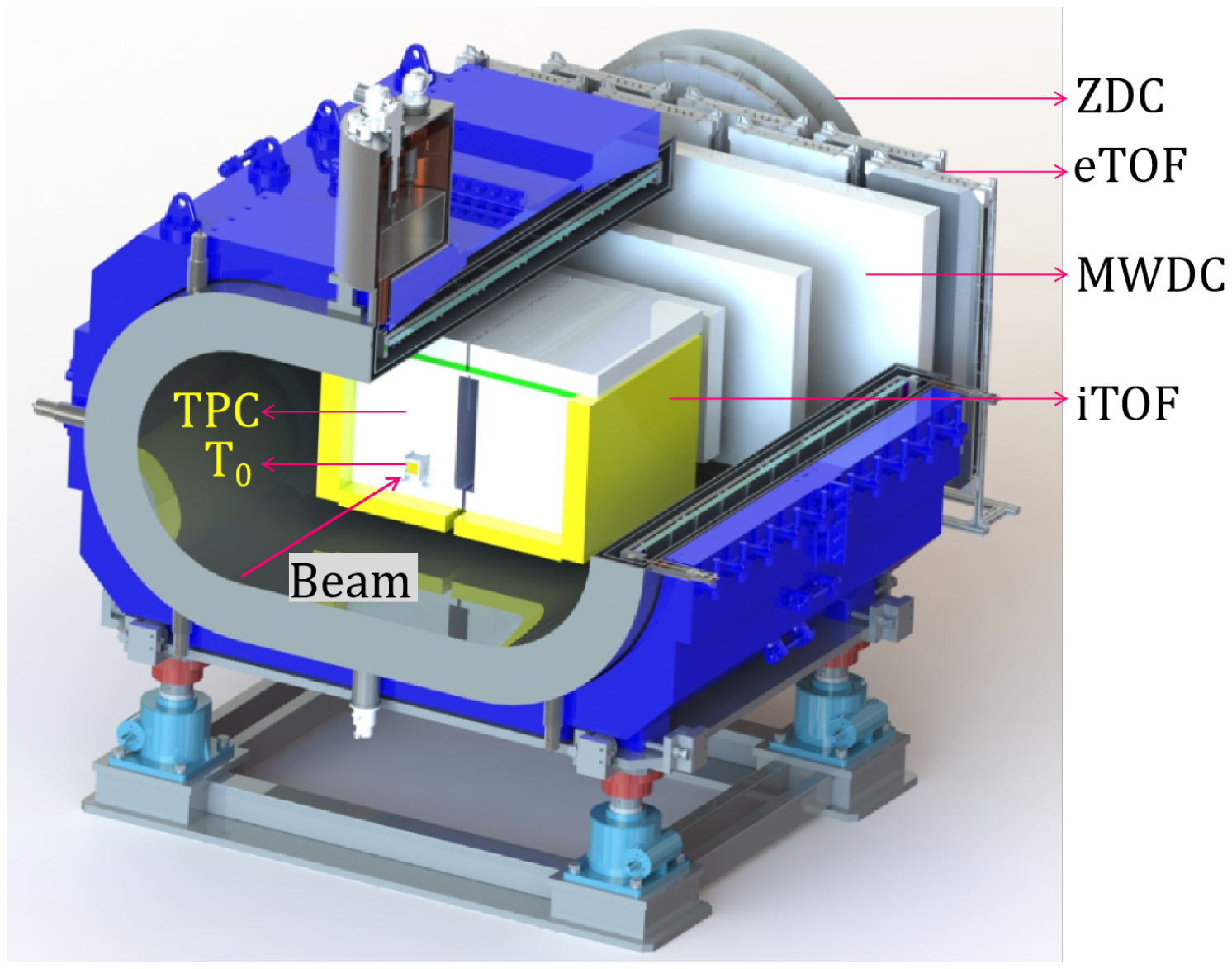}
\caption{(Color online) The technique design of the CSR external-target experiment (CEE). Details of AC and BM are not shown.}
\label{cee}
\end{figure}

With a decade of researches and developments (R\&D), the CSR external-target experiment (CEE) is currently in construction ~\cite{YH2014,LLM2016,HD2017}. It has the ability to measure the light charged particles (LCPs) covering almost the whole $4\pi$ space in center of mass reference.  Fig. \ref{cee} presents the schematic view of the CEE experiment.  The main component of the CEE is a large-gap magnetic dipole, housing the tracking detectors. For the midrapidity, the LCPs are tracked by two time projection chambers (TPCs) placed side-by-side with an empty space in the middle for the beam passing through. On the left, right and bottom sides of the TPCs, an inner time-of-flight wall (iTOF) is mounted to measure the TOF information of the tracks at midrapidity. For the large rapidity region, an array consisting of three multi wire drift chambers (MWDCs) followed by an external TOF wall (eTOF) is  designed for measuring the LCPs. Both eTOF and iTOF are made of multi-layer resistance plate counters (MRPC), requiring the time resolution of about 60 and 50 ps, respectively.  A zero-degree counter (ZDC) is mounted on the most downstream side to determine the reaction plane as well as to provide complementary information to derive the centrality. The other sub detectors, including the start time detector $\rm T_0$, the active collimator (AC) and the silicon pixel beam spot monitor (BM) are all mounted on the upstream side of the target on the beam line. 

Since the start of the CEE project, enormous R\&D have been conducted. All the sub-detectors have been proceeded over the design stage. Currently, all prototypes of  CEE  have been tested, using cosmic rays and ion beams, and entered the stage of engineering verification ~\cite{WBT2020,WBT2023, HD2020,ZSH2021,WX2022,WHL2022,LJ2023, GD20241,GD20242,HZB2024}. Particularly for the MWDC, the design, manufacture, assembly and source test have been presented in \cite{HZB2024}. In this paper, we focus on the beam test results of the MWDC prototype and demonstrate that its  performance meets the requirements of the design of CEE. The remainder of the paper is arranged as following. Section 2 presents the experimental setup of the beam test.  The results and discussions are presented in section  3. Section 4 is the conclusion.

\section  {Experimental setup of the beam test}    \label{sec. II}  

\begin{figure}[htb]
\centering
\hspace{-0.7cm}
\includegraphics[width=0.5\textwidth]{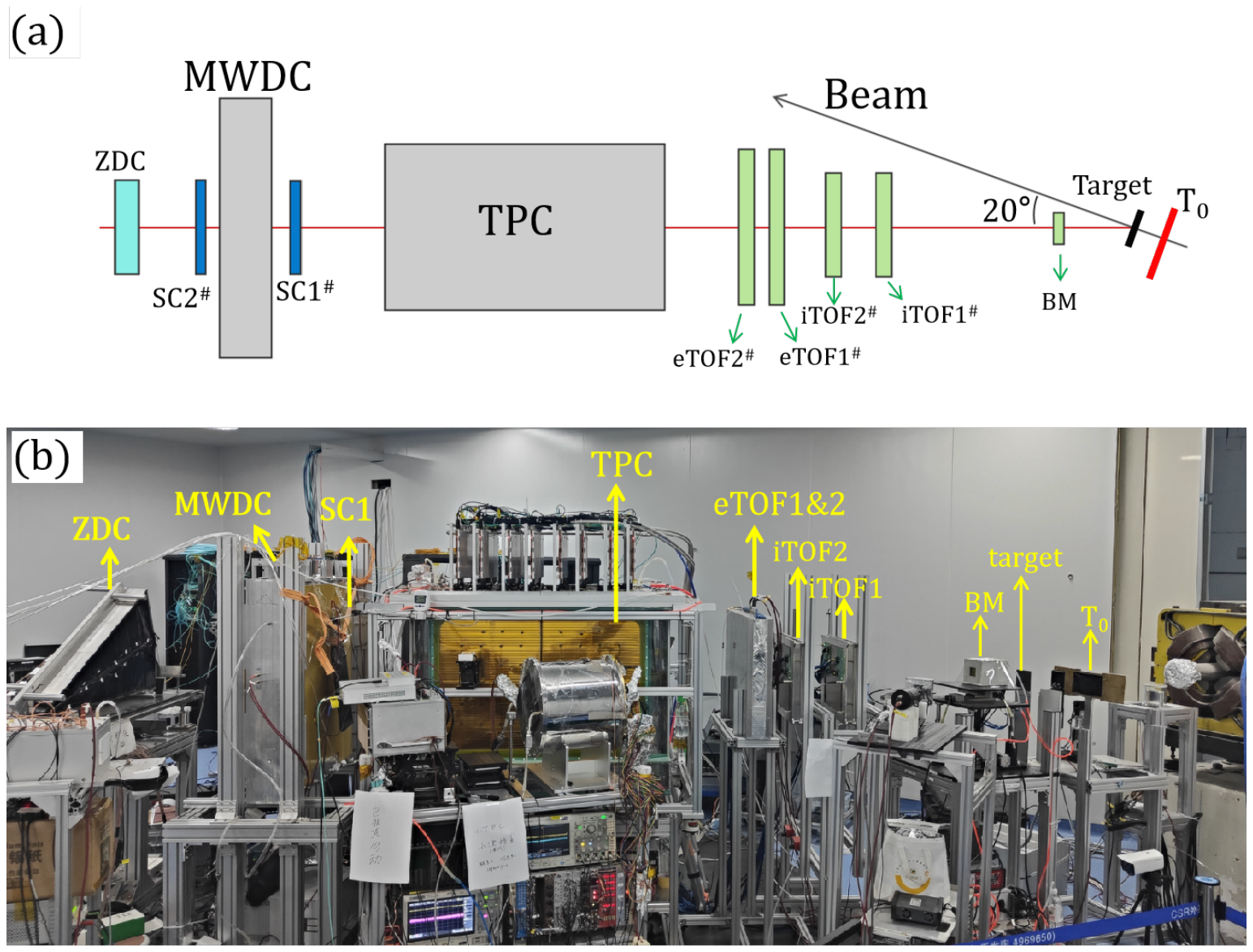}
\caption{(Color online) The detector setup in the test beam experiment performed at the CEE location on HIRFL-CSR. (a) a schematic view, (b) a real photograph.}
\label{setup}
\end{figure}

 The beam test experiment was performed on the site where the CEE will be located at HIRFL-CSR. The 350 MeV/u Kr beam was delivered by the main ring of the CSR, hitting an iron target in the thickness of about 1 mm. Fig. \ref{setup} presents the schematic view (a) and a real photograph (b) of the detector setup in the beam test.  The prototypes of the sub detectors  are installed at about $20 ^\circ$ with respect to the beam direction. Behind  the target is the start timing detector $\rm T_0$, followed by two iTOF MRPCs iTOF1 and iTOF2, and two eTOF MRPCs eTOF1 and eTOF2, respectively. Then the full-size TPC prototype  follows. The MWDC, sandwiched by two scintillators SC1 and SC2, are placed about 20 cm behind the TPC. On the most downstream of the setup, is the prototype of ZDC.  

\begin{figure}[htb]
\centering
\hspace{-0.7cm}
\includegraphics[width=0.5\textwidth]{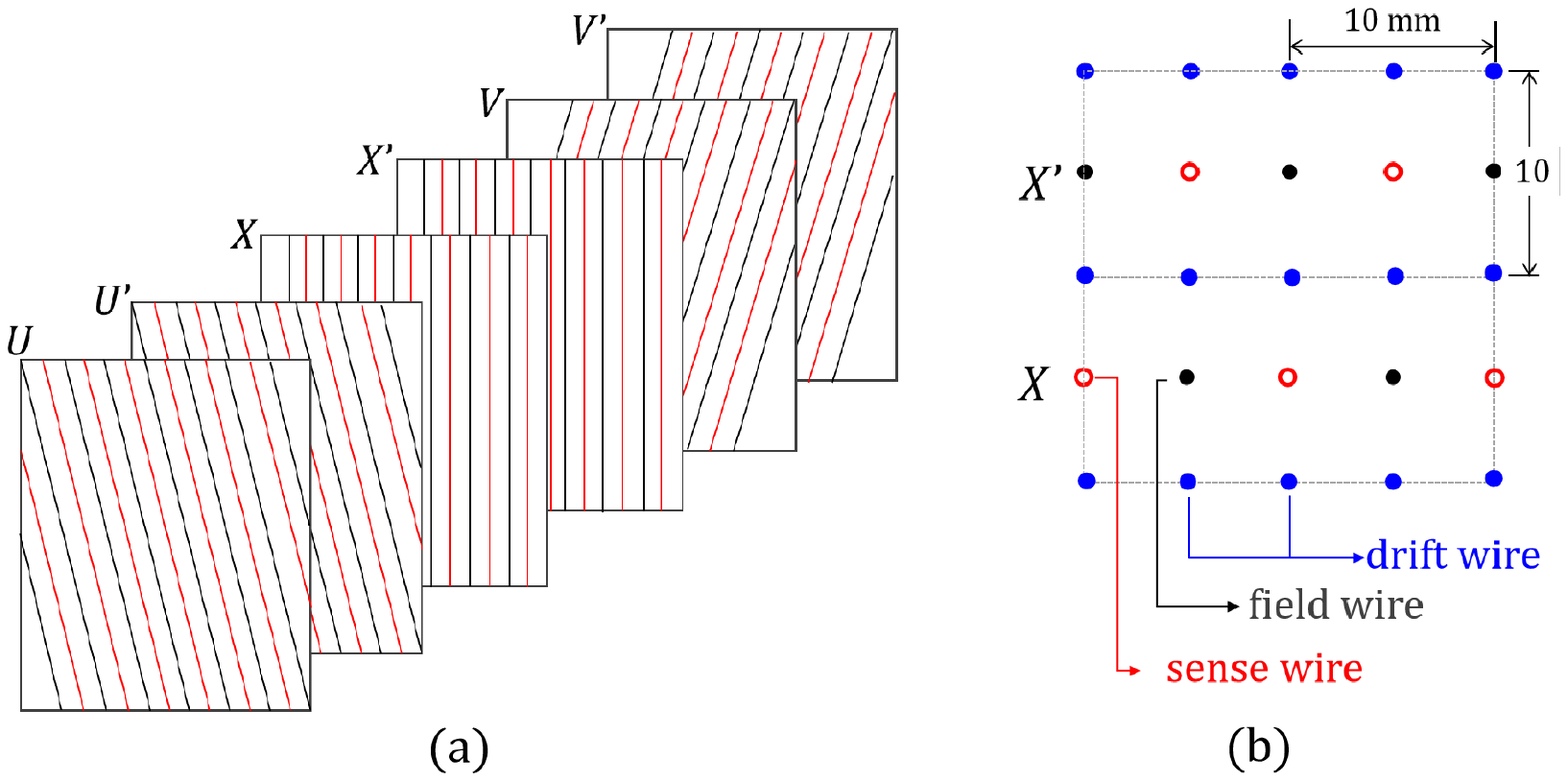}
\caption{(Color online) (a) The schematic view of the sense wire orientation, (b) The cross section view of the drift cells along X sense wires.}
\label{MWDC}
\end{figure}

The half-size prototype of the MWDC  has the same structure with the real MWDC to be mounted on CEE. But the sensitive area of the prototype is $\rm 76 \times 76 ~cm^2$, which is smaller than that of the real one. The prototype  consists of 6 sense wire layers, The sense wires are stretched in three directions, namely  X, U and V meeting $0^\circ$, $30^\circ$ and $-30^\circ$ with respect to the vertical axis, respectively, as shown in Fig. \ref{MWDC} (a).  The size of each drift cell is $\rm 10 mm \times 10 mm$, i.e., the inter sense wire distance is 10 mm.  As an example, Fig. \ref{MWDC} (b) presents schematically the  wire distributions  of two neighbouring drift cells of the X sense plane, from the point of view of the transverse cross section. On each direction, the two parallel sense planes are displaced by half cell in order to remove the left-right ambiguity when a track passing through. The high voltage (HV) fed to the sense wire is 1500 V, and the working gas is the mixture of 80\% argon and 20 \% $\rm CO_2$ slightly beyond the atmosphere pressure. Under this working condition, the energy resolution  is about 23\% for the 5.9 keV X ray in $\rm ^{55}Fe$ source test \cite{HZB2024}.

\begin{figure}[htb]
\centering
\hspace{-0.7cm}
\includegraphics[width=0.5\textwidth]{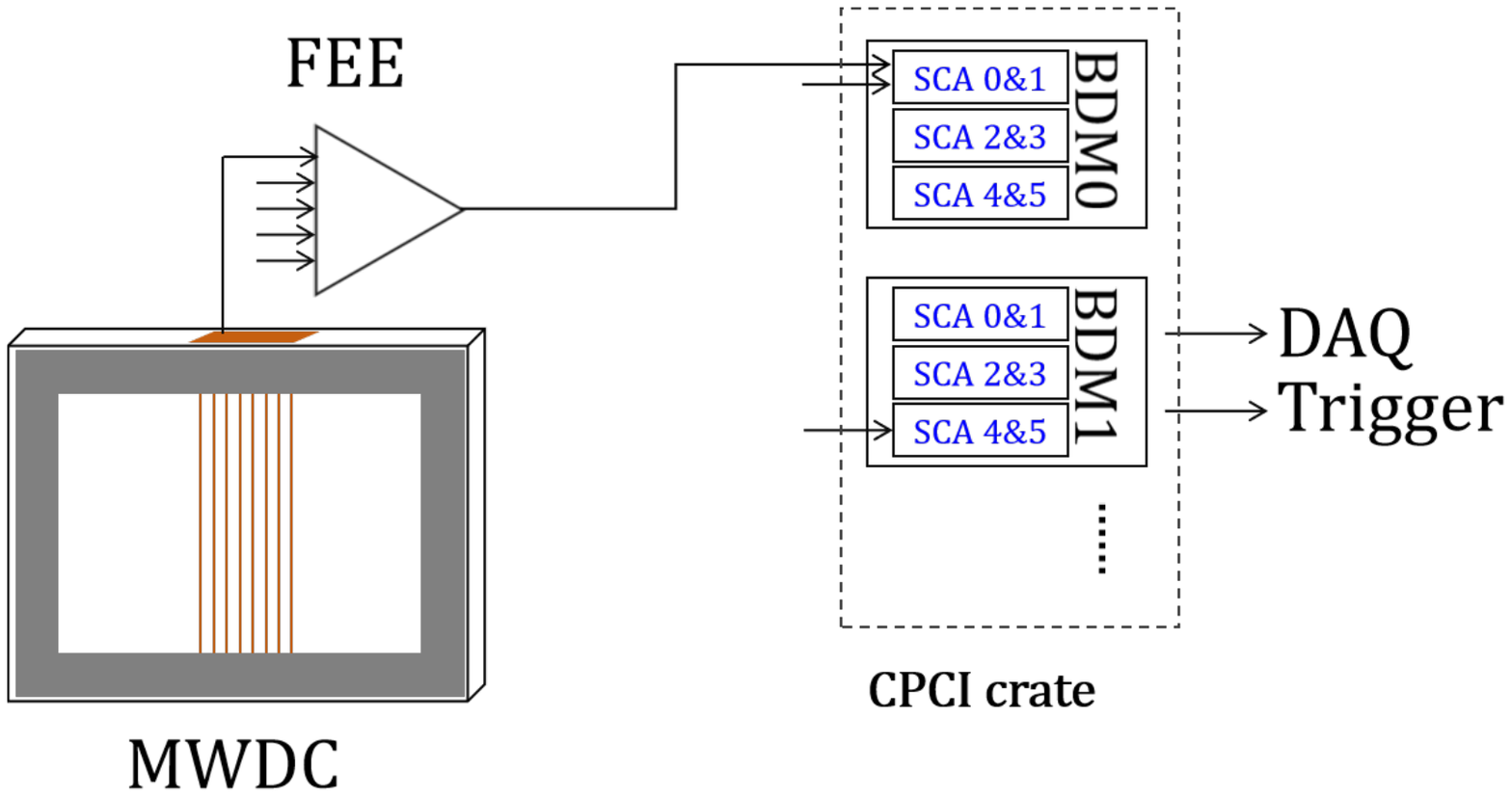}
\caption{(Color online) The schematic view of the electronics and the signal flow of the MWDC prototype.}
\label{electronics}
\end{figure}

The signals from the sense wire are amplified and shaped in the  front-end electronics (FEE) mounted on PCB boards, with each one housing 16 channels. The FEE signals are then sent to a serial capacity array (SCA) for the digitization \cite{Hao2019,Xu2022,Liu2021}, where the pulse  are sampled by a 12.5 MHz frequency followed by the analog to digit conversion. Each SCA board houses 16 channels, and 6 SCA boards are grouped on one back-end data module (BDM), which reads out the data from the SCA and  further sends them to the main data acquisition (DAQ) system if a global trigger signal arrives. In the beam test, totally 192 channels of the MWDC prototype are connected, corresponding 32 wires which are readout on each sense wire plane.   Fig. \ref{electronics} presents the electronics scheme and the signal flow of the MWDC prototype.

The trigger system was built with field programmable gate array (FPGA) technology. The signals from the iTOF and eTOF detectors are collected by the FEE boards to calculated the multiplicity, which are fed to the main trigger module (MTM) via optical fiber connections. If the trigger condition is met, the  MTM generates a trigger signal and sends it to all sub detectors via downlink chains. In the beam test, our trigger condition requires the coincidence of the $\rm T_0$, iTOF and eTOF. For the detailed construction of the FPGA-based trigger system, one can refer to \cite{GD20242}. 


\section {Results and Discussions}   \label{sec. III}

In this section, we present the performance of the MWDC prototype in the beam test experiment. 

\begin{figure*}[htb]
\centering
\hspace{-0.7cm}
\includegraphics[width=0.75\textwidth]{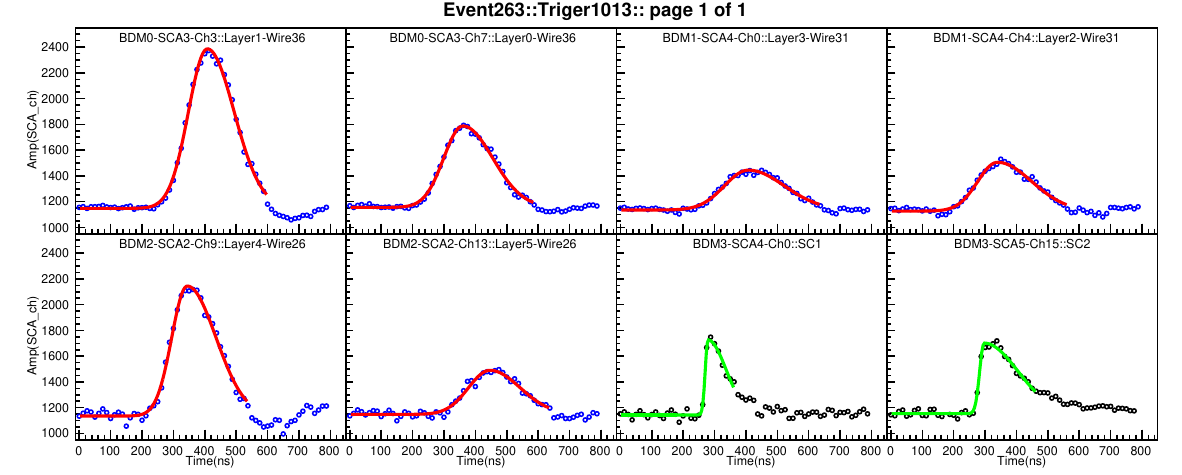}
\caption{(Color online) An event display with 6 wires fired. The last two panels in the lower row present the signals from SC1 and SC2, respectively. The pulses are fitted using asymmetric Gaussian function.}
\label{event_display}
\end{figure*}

Fig. \ref{event_display} presents a typical event display, where one can find 6 wires being fired by an incident particle. The last two panels with green fit denote the signals of the the two scintillators SC1 and SC2, respectively.   The abscissa is the sampling times in the unit of nanosecond. The ordinate is the amplitude of the signal. The signal shape is smooth, with a rising time of about 100 ns and a descent time of about 200 ns.  Second, as seen from figure, the base line is quite stable in each individual channel. The fluctuation of the base line for a single channel is about 25 channels, corresponding to 8 mV according to our calibration. 

\begin{figure}[htb]
\centering
\hspace{-0.7cm}
\includegraphics[width=0.45\textwidth]{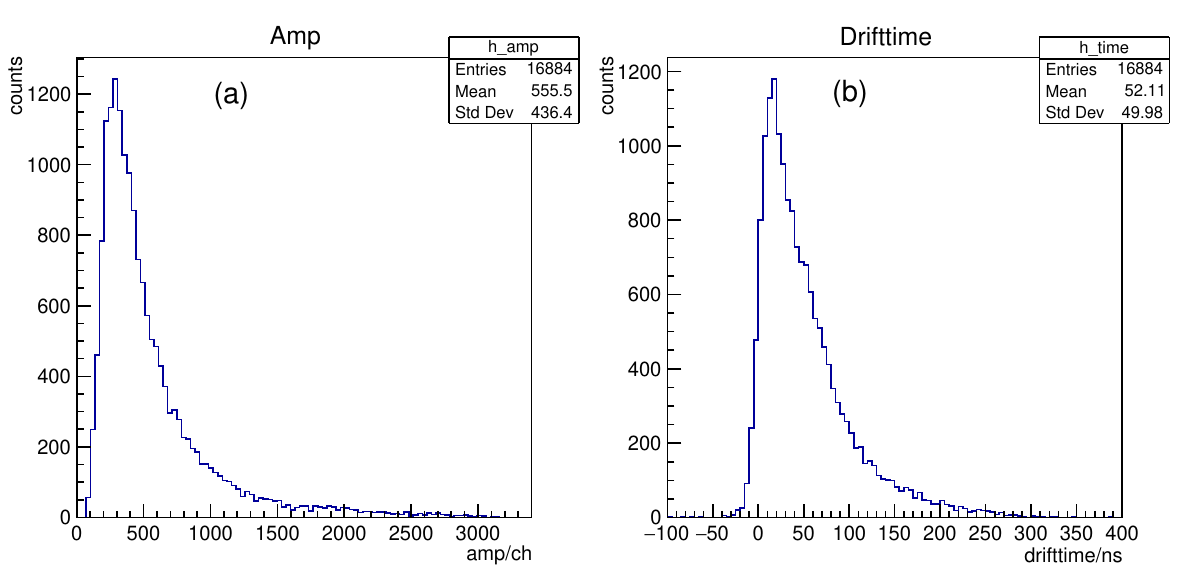}
\caption{(Color online) The distribution of the energy loss (a) and drift time with respect to the timing provided by SC1 and SC2 (b).}
\label{signal}
\end{figure}

Once a signal is identified for an individual channel,  the amplitude is calculated by the integration of the peak with the baseline being subtracted.   Fig. \ref{signal}  (a) presents the distribution of the signal amplitude from the fired wires. It is shown that the distribution exhibits approximately the Landau distribution as expected. The most probable value of the amplitude situates at about 200 channel, corresponding to about 70 mV. Assuming the incident particles are mainly protons with an average kinetic energy of 200 MeV, then one can estimate the most probable energy loss is about 5.4 keV. The energy deposit of the minimum ionization particles is averagely 2 keV, and is expected to situate in the vicinity of 25 mV, which is beyond the baseline fluctuation.  

For the same signal, the rising and falling edge is fitted using a asymmetric Gaussian function, and the timing information is extracted from the constant fraction of 0.1. 
Fig. \ref{signal} (b) presents the distribution of the drift time for the fired wires with respect to the timing provided by SC1 and SC2. It is shown that except for the main peak, with a width of about 100 ns, there is also a long tail corresponding to the signal originating from the corners of the drift cell. The main peak rises very rapidly, indicating a rather good zero drift time determination.  

\begin{figure}[htb]
\centering
\hspace{-0.7cm}
\includegraphics[width=0.45\textwidth]{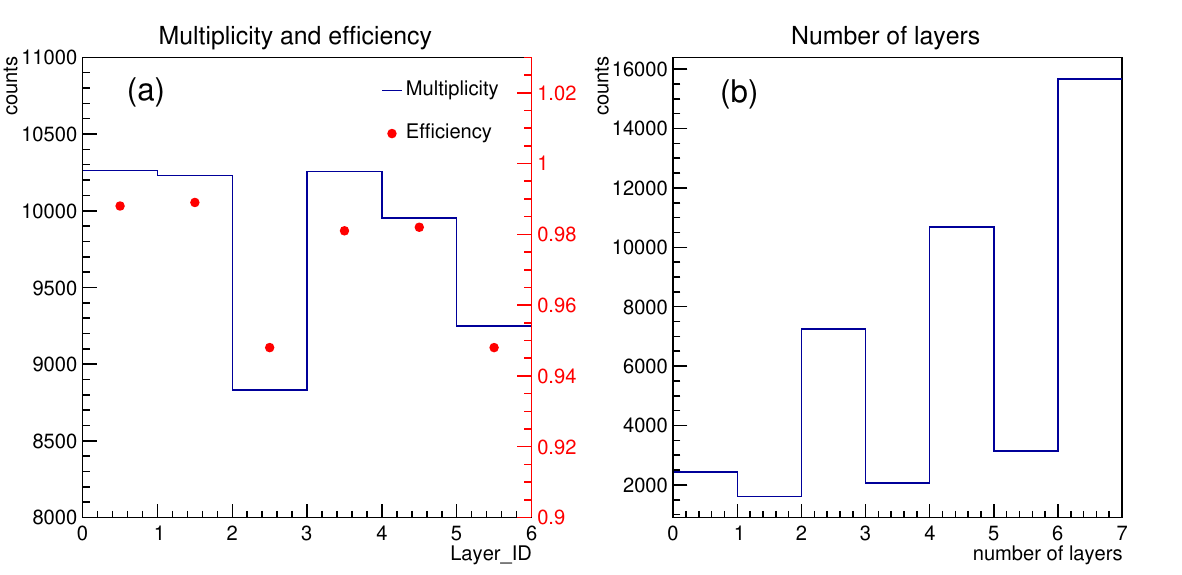}
\caption{(Color online)  (a) The distribution of the multiplicity of the fired layers. (b) The efficiency of each layer at 1500 V working voltage. }
\label{efficiency}
\end{figure}

The histogram in  Fig. \ref{efficiency} (a) presents the count distribution of all the six layers  for the events triggered by the  coincidence of the two scintillators SC1 and SC2.  It is seen that for the layer 2 and 5, the counts are slightly  low because there is one dead channel in each of these two layers. The efficiency of the 6 layers are represented by the solid dots shown on the right ordinate. The efficiency is 95\% for  layer 2 and 5 and 98\% for others.   Shown in  panel (b) is  the distribution of the multiplicity of the fired layers. Clearly, most of the events fired even layers. This is expected because the fired wires for a given direction in $X$, $U$ and $V$ usually come together in pair. Since the coverage of the connected wires are slightly small than the area of the scintillators in front of and behind the MWDC prototype, there are some tracks missing the connected wires, causing the multiplicity distributing at  2 or 4 partially.

\begin{figure}[htb]
\centering
\hspace{-0.7cm}
\includegraphics[width=0.5\textwidth]{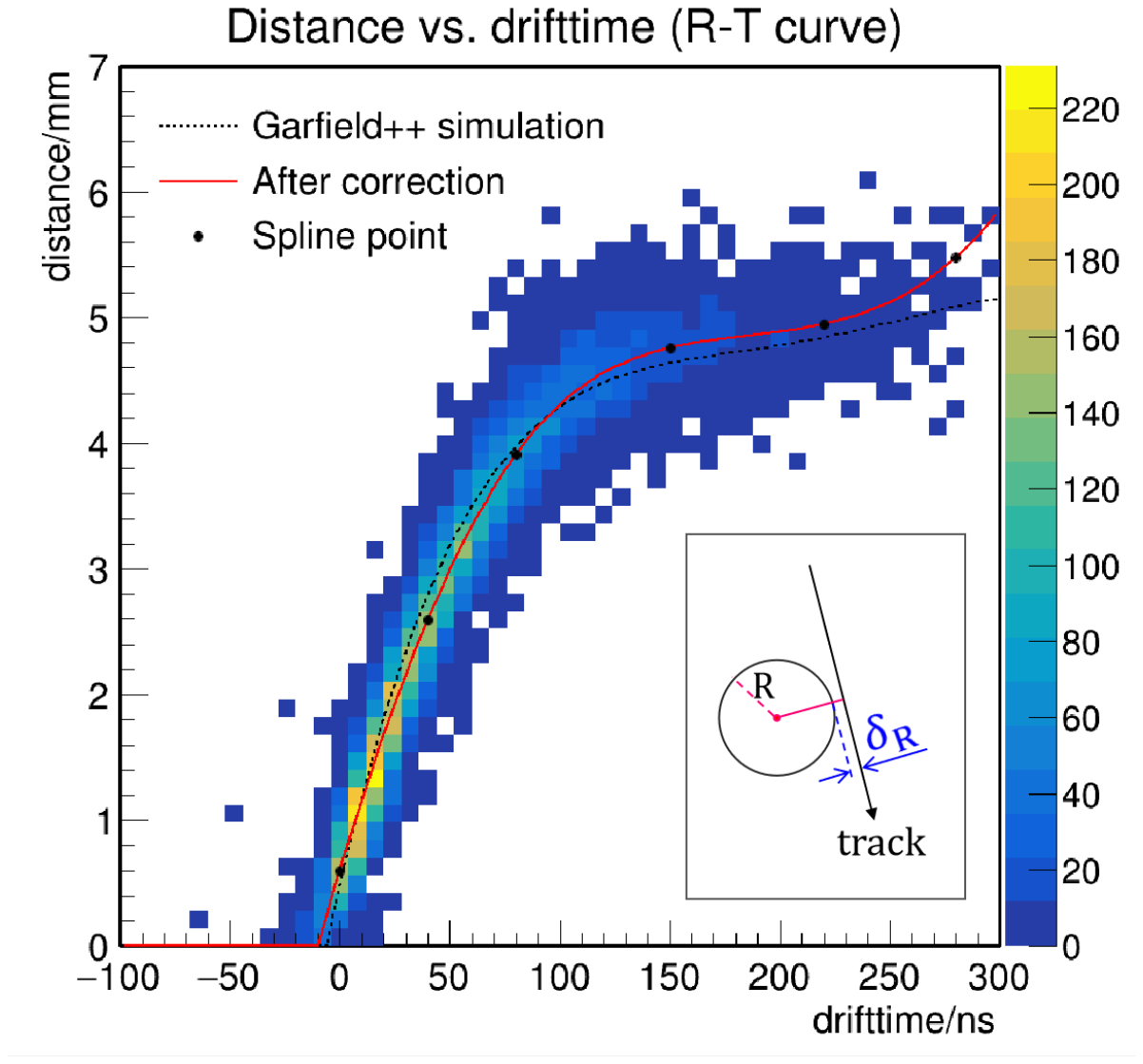}
\caption{(Color online)  The $R\text{-}T$ curve calibration. The inset shows the definition of the tracking residue $\delta_{\rm R}$.  }
\label{rt}
\end{figure}

Before showing the  tracking performance of the MWDC prototype, we investigate the $R\text{-}T$ calibration, where $R$ and $T$ are the drift distance and drift time, respectively. The calibration has been done in an iteration scheme. The calculated curve from {\small \ttfamily Garfield++ } code \cite{garfield} is adopted as the initial $R\text{-}T$ curve, as shown by the dashed curve in Fig. \ref{rt}.  With this initial condition, the tracks are fitted and the tracking residue  is calculated as a function of the drift distance $R$. Once the tracking residual has been obtained, it is used to correct the drift distance and the fit is then redone. After several iterations, the $R\text{-}T$ curve is converged. The solid curve represents the calibrated $R\text{-}T$ curve. From the curve it can be derived that the drift velocity at the middle of the drift cell is about  $v_{\rm d}\approx 4.5~ {\rm cm/\mu s}$.  The color palette presents the $R\text{-}T$ distribution for all the events. The width of this distribution contains the information of the tracking residue  $\delta_{\rm R}$, which  is defined as the difference between the measured drift distance $R=Tv_{\rm d}$ and the expected drift distance $R_{\rm exp}$ after tracking, i.e., $\delta_{\rm R}=R-R_{\rm exp}$, as depicted in the inset of Fig. \ref{rt}.  

\begin{figure}[htb]
\centering
\hspace{-0.7cm}
\includegraphics[width=0.45\textwidth]{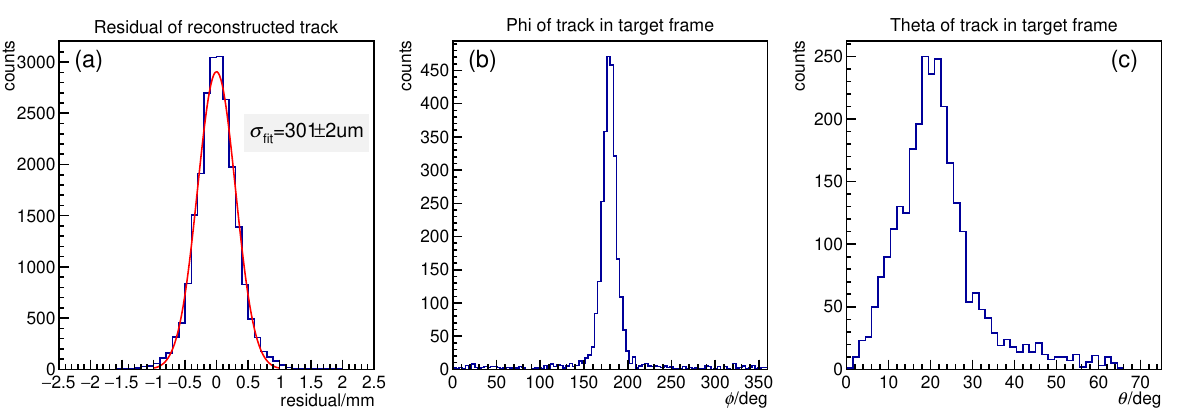}
\caption{(Color online)  The distribution of the tracking residue. The curve is the fitting using Gaussian function. }
\label{residue}
\end{figure}

Fig. \ref{residue} (a) presents the tracking residue distribution. Here we only count the tracks firing 6 layers of sense wire planes. It is shown that the distribution exhibits a Gaussian shape.  Using a  Gaussian fit, the residue is derived $\sigma_{\rm t}=301\pm2 {\rm \mu m}$. This result is  consistent with our result of a $\rm 10~cm\times 10~cm$ prototype tested using cosmic ray muons \cite{YH2014}. It suggests that the tracking performance is not deteriorated when the sensitive area becomes larger. The tracking residue is consistent to the expected performance of CEE MWDC array, which is $300 ~{\rm  \mu m}$ in the design. Defining the beam direction as the $Z$ axis and the vertical direction as $Y$ axis in laboratory, one can get  the azimuth angle $\phi$ and the polar angle $\theta$ distribution for the tracks measured by the MWDC prototype, as shown in panel (b) and (c), respectively. The peaks of  $\theta$ and $\phi$ situate at $\left<\theta\right>=20^\circ$ and $\left<\phi\right>=180^\circ$, indicating the tracks are originating from the target position as expected.

 \begin{figure}[htb]
\centering
\hspace{-0.7cm}
\includegraphics[width=0.5\textwidth]{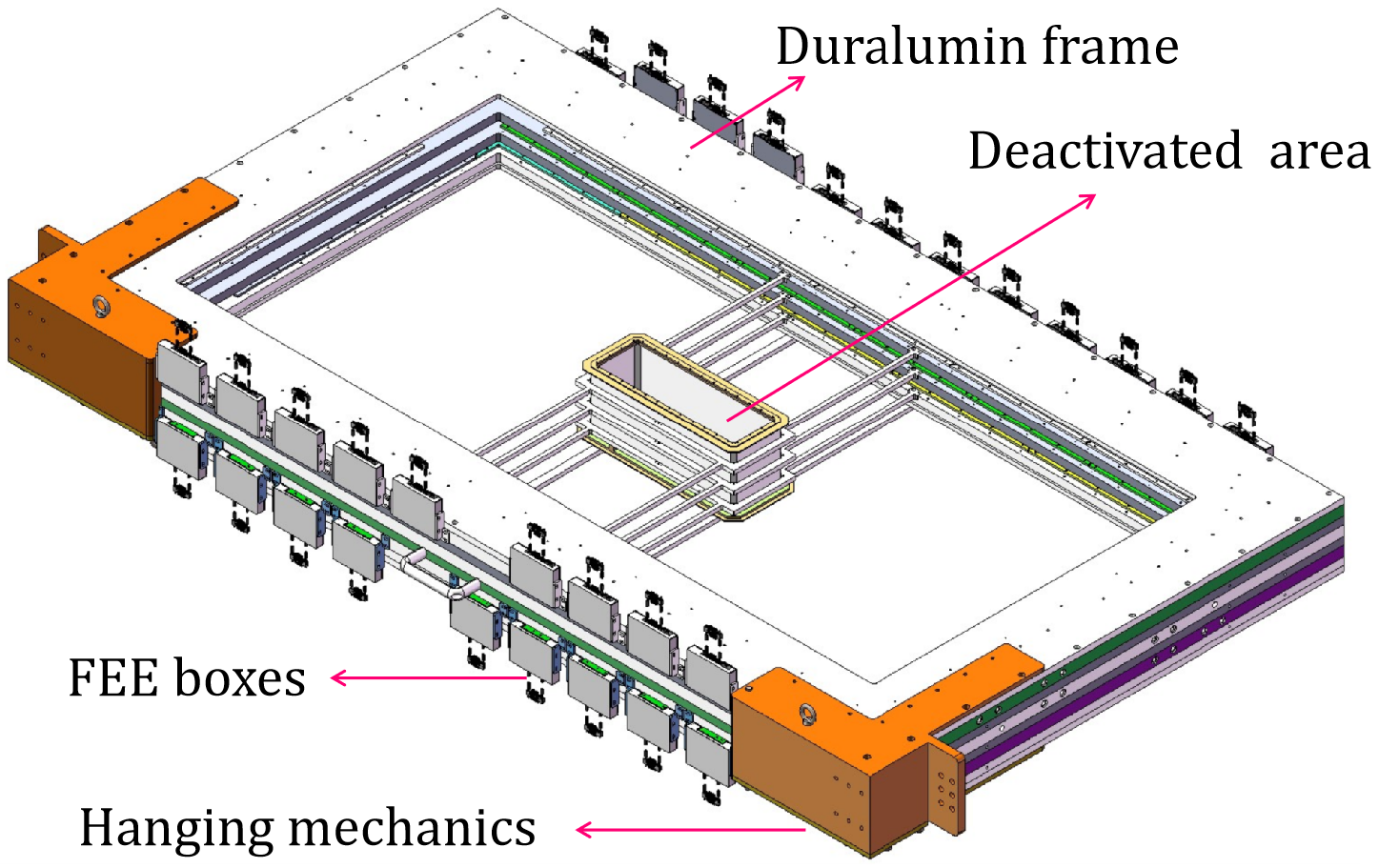}
\caption{(Color online) The mechanic structure of the MWDC. In the center is the inactive area, through which the unreacted beam particles pass through. The sensitive area of the detector is $\rm 136~cm (x) \times 63~cm (y)$  and the deactivated area is  $\rm  30~cm~(x) \times  10~cm~(y)$.}
\label{inactive}
\end{figure}

 The final  problem remains to be solved for the application of the MWDC. In real beam experiment, the unreacted  beam particles, which are the majority, pass through the MWDC array of CEE and cause serious saturation problem to the whole system. In order to cure the malfunction arisen from the unreacted beam, we design an deactivated area for each MWDC.  Fig. ~\ref{inactive} presents the engineering prototype of the MWDC, which has  been assembled recently and has the same size as the real MWDC1 to be mounted on CEE.   It is shown a stack of hollow rectangle PCBs, with each one corresponding to each layer of the wire frame, are mounted in the center of each MWDC. For the wires in the vicinity of the beam line, in stead of going through the whole area, these wires are  soldered to the PCB as a bridge circumventing the beam path. The size of the deactivated  area is about $\rm  30~cm~(x) \times  10~cm~(y)$  for the nearest MWDC  and $\rm  60~cm~(x) \times  ~20cm~(y)$ for the furthest MWDC, respectively.  To verify that the design works, a small MWDC prototype (sensitive area $\rm 18~cm \times 18~cm$) with such a structure is mounted on the beam line during the test experiment.  At an beam intensity of $2\times10^6$ pps, which is higher than the usual condition in future experiment, the rate of the signals with large amplitude on the MWDC is much less than $10^{-3}$ of the beam intensity. And these signals are mainly induced by the projectile-like products of the interaction between the beam and the upstream material on the path,  and bring no problems to the operations.

\begin{figure}[htb]
\centering
\hspace{-0.7cm}
\includegraphics[width=0.45\textwidth]{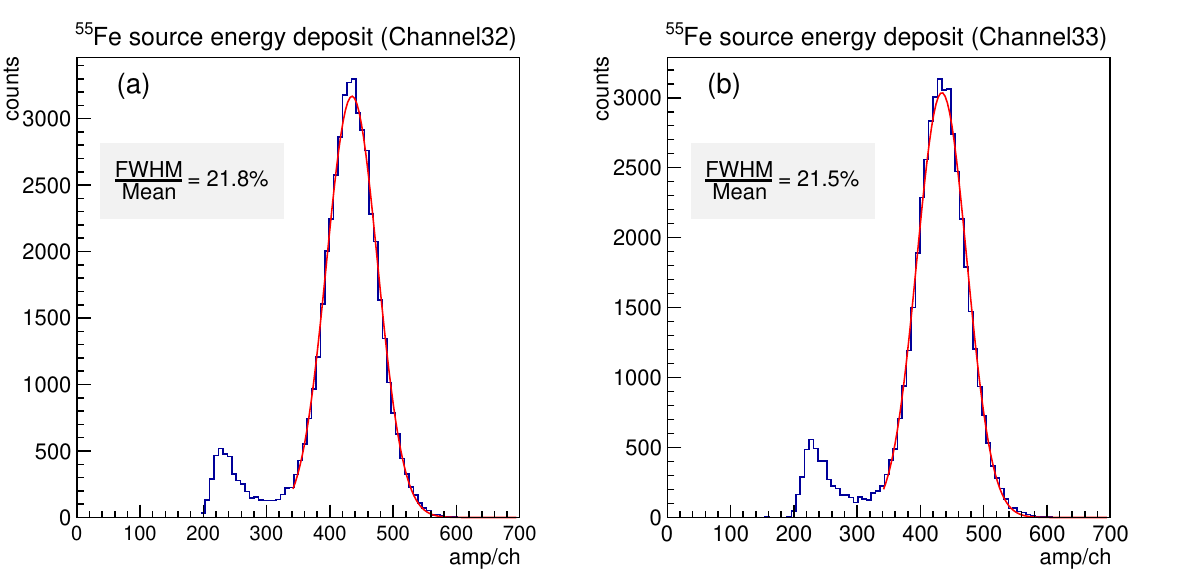}
\caption{(Color online) The  energy spectra of the X-rays  from the $^{55}$Fe source in two channels  for the engineering  prototype shown in Fig. \ref{inactive}. The operation HV here is 1400 V.  }
\label{iron}
\end{figure}

It is also checked that the introduction of the deactivated area does not influence of the energy resolution of the engineering prototype detector. As an example, Fig. \ref{iron} presents the energy spectra of the X-ray of  $^{55}$Fe source recorded in two  neighbouring wires from the engineering prototype. The electronics setup is the same with Fig. \ref{electronics}. The results show that the energy resolutions is about $22\%$, in accordance with the prototype without the deactivated area \cite{HZB2024}.  With  the results of the beam test and the source test, it is  demonstrated that  the prototype of the MWDC, together with the electronics, can meet the requirements of the CEE experiment.

\section{Conclusion}  \label{sec. IV}

To conclude, the prototype of the MWDC detector at CEE has been assembled and operated in a beam test experiment of 350 MeV/u Kr+Fe.  The prototypes of MWDC and all other detectors, the trigger systems and the DAQ systems  are  integrated in test beam experiment. Being operated at 1500 V working voltage, the MWDC works correctly with an average efficiency beyond 95\% for all the sense wire layers. The baseline fluctuation is about 7 mV. The $R-T$ curve has been calibrated and the drift velocity of $v_{\rm d}\approx 4.5 ~{\rm cm/\mu s}$ has been obtained. For the events with $M_{\rm layer}=6$, the tracking residue is about  $301 \pm 2 \rm \mu m$ . The energy resolution of the engineering prototype is typically 22\% for the $\rm ^{55}Fe$ X-rays, without being influenced by the introduction of a deactivated area in the center to avoid the malfunction caused by the unreacted  beam particles. The experimental results demonstrate that the MWDC meets the requirement of the CEE experiment.

\end{document}